# Secular Brightness Trends of Starlink and OneWeb Satellites

Anthony Mallama [1]

[1] IAU Centre for the Protection of the Dark and Quiet Sky,
IAU-UAI Headquarters, 98-bis Blvd Arago, 75014,
Paris, France

2026 April 14

Correspondence: anthony.mallama@gmail.com

Magnitudes recorded from 2021 through 2026 for Starlink VisorSats reveal brightening of 0.6 magnitudes, while those of OneWeb show dimming of 0.4. Both trends are significant at 3 sigma. This study is based on 1.6 million magnitudes recorded by the MMT9 robotic observatory.

**1. Introduction**

Bright spacecraft interfere with astronomical observations and aesthetic appreciation of the night sky (Barentine et al. 2023, and Mallama and Young 2021). In order to assess the severity of this problem, luminosity characteristics of all the satellite constellations have been reported by Mallama and Cole (2025) and Mallama et al. (2025). Those studies were usually conducted soon after the first satellites of a constellation were launched. This study reports on long-term, multi-year monitoring studies that reveal brightness changes for satellites at operational altitudes for the two oldest spacecraft constellations, Starlink and OneWeb.

Section 2 describes the observations and the data processing. Section 3 characterizes the brightness changes for Starlink VisorSat and V1.5 satellites, while Section 4 characterizes OneWeb spacecraft. Section 5 discusses the results.

**2. Observations**

Photometric measurements were recorded by the MMT9 robotic observatory in Russia (Karpov et al. 2015 and Beskin et al. 2017) located at 43.65N and 41.43E. The magnitudes are within 0.1 of the Johnson V-band as discussed by Mallama (2021).

MMT9 records the data in tracks across the sky at a frequency of 10 Hz. We selected tracks obtained when satellites were at their operational heights. The mean apparent magnitude for each track was computed. We also took means of magnitudes adjusted to a distance of 1,000 km (according to the inverse square law of light) for greater uniformity. Those two means for every track were then binned into time segments of 0.1 year duration for further analyses. Phase function corrections were not applied because they add an element of uncertainty. This study centers on the 1000-km magnitudes.

### 3. Starlink spacecraft

VisorSats were an early attempt by SpaceX to mitigate satellite brightness by shading the nadir facing chassis from sunlight. The first launch with all VisorSat spacecraft occurred on 2020 August 7.

We analyzed 317,224 VisorSat magnitudes for 34 satellites obtained during 858 tracks. Figure 1 shows that their brightness increased between 2021 and 2026. The rate of change for 1,000-km magnitudes is -0.12 magnitude per year (based on a linear least squares fit) and that change is statistically significant at 3.1 sigma, as indicated in Table 1.

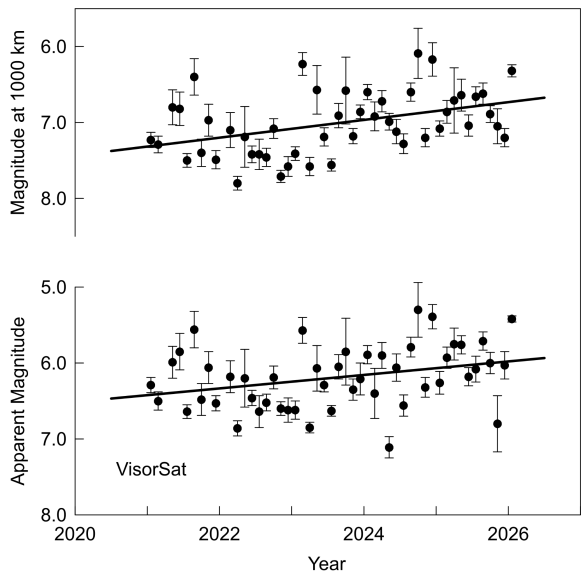

Figure 1. The brightening trend for VisorSat.

SpaceX omitted the visors from their V1.5 spacecraft because they interfered with laser communications. We analyzed 236,058 V1.5 magnitudes for 35 satellites recorded during 640 passes. Figure 2 suggests a brightness increase between 2022 and 2026. However, the rate of change for 1,000-km of -0.07 magnitude per year is not statistically significant.

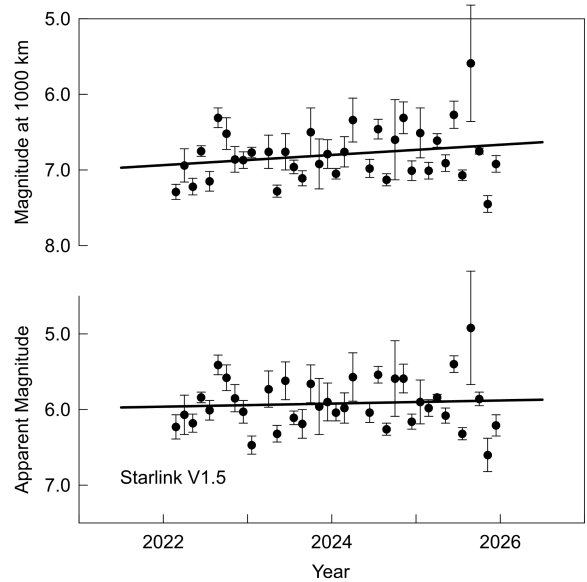

Figure 2. The brightening trend for Starlink V1.5.

### 4. OneWeb satellites

Eutelsat operates the OneWeb constellation of satellites. We analyzed 1,069,435 magnitudes of 35 satellites recorded during 1,451 tracks.

OneWeb satellites faded between 2021 and 2026 as shown in Figure 3. The dimming of 0.078 magnitudes per year for 1,000-km magnitudes is significant at 3.9 sigma.

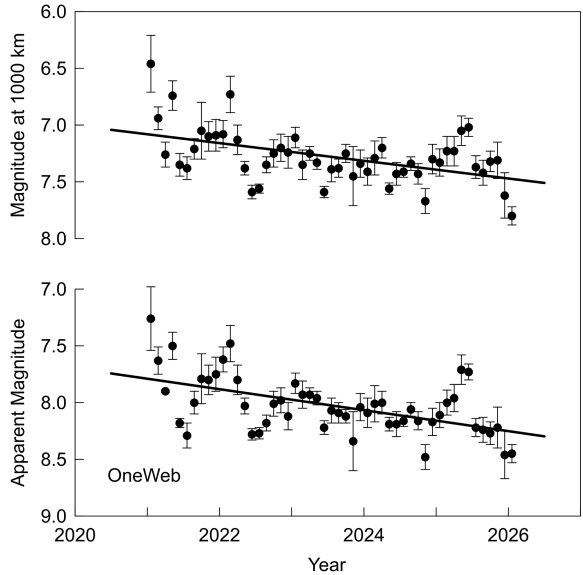

*Figure 3. The dimming trend for OneWeb.*

Table 1. Brightness trends

```
              Slope    StDev   Sigma
VisorSat
1000-km      -0.122    0.038    3.1
Apparent     -0.089    0.039    2.3

Starlink V1.5
1000-km      -0.067    0.051    1.3
Apparent     -0.020    0.049    0.4

OneWeb
1000-km      +0.078    0.020    3.9
Apparent     +0.092    0.021    4.4
```

**5. Discussion of brightness trends**

The fading of OneWeb satellites does not appear to be due to aging. In order to investigate the effects of materials aging and space weathering we studied data for 'later' spacecraft launched on 2024 October 20 separately from the 'early' ones launched prior to 2022. The early satellites were characterized in Section 4.

We analyzed 61,088 magnitude of 19 later satellites collected during 95 tracks. Their average 1000-km magnitude during the period 2025.25 through 2026.05 was 7.41 +/- 0.16. This is dimmer than the average of early satellites in the same period (7.35 +/- 0.08) although the difference is not significant.

The early and later satellites are plotted together in Figure 4. Their similar brightness suggests that material aging and space weather are not causing the OneWeb satellites to fade.

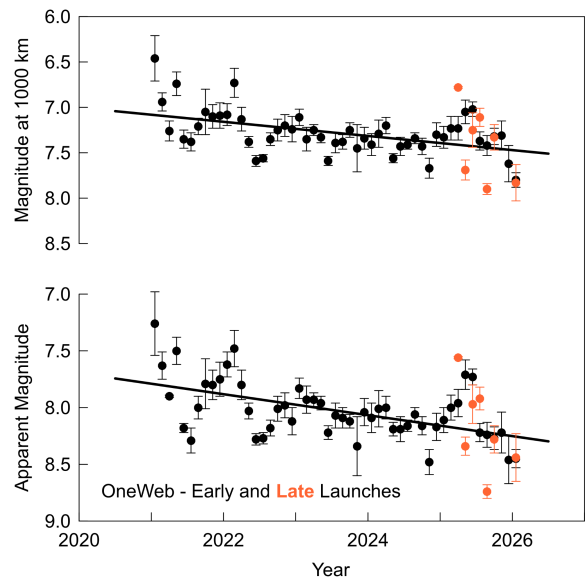

*Figure 4. OneWeb satellites from early and later launches were nearly the same brightness in 2025-2026.*

Starlink VisorSat and V1.5 satellites were discontinued when SpaceX began launching V2 spacecraft. So, we could not compare the brightness of early and late spacecraft.

The root cause of the observed brightness changes is not known. Communications with SpaceX and Eutelsat did not resolve these issues.

## 6. Conclusions

OneWeb satellite became significantly fainter during a five year period at a rate of 0.078 magnitudes per year, while Starlink VisorSat spacecraft brightened by 0.122 magnitudes per year. Secular studies of satellite luminosities are needed in order to assess their brightness trends.

## 7. Data and software availability

Magnitudes for the satellites studied here are available from the MMT9 online database at http://mmt9.ru/satellites/. Software used in this research is available from the author.

**Acknowledgements**

M. Dickinson conducted the review of this paper for the IAU CPS. Two reviewers provided helpful suggestions.

R.E. Cole created the PowerShell file used to download batches of MMT9 observations.

S. Karpov provided information about MMT9 satellite observations and E. Katkova maintains the public database of MMT9 magnitudes.

The authors acknowledge the support of the International Astronomical Union (IAU) Centre for the Protection of the Dark and Quiet Sky (CPS). The Centre coordinates collaborative and multidisciplinary international efforts from institutions and individuals working across multiple geographic areas, seeks to raise awareness, and mitigate the negative impact of satellite constellations on ground-based optical, infrared and radio astronomy observations as well as on humanity's enjoyment of the night sky. Any opinions, findings, and conclusions or recommendations expressed in this material are those of the author(s) and do not necessarily reflect the views of the IAU, NSF NOIRLab, SKAO, ESO, or any host or member institution of the IAU CPS.